\documentclass[prl,twocolumn, preprintnumbers,amsmath,amssymb,superscriptaddress]{revtex4}

\usepackage{graphicx}
\usepackage{dcolumn}
\usepackage{bm}
\usepackage{float}

\begin{document}

\title{Tunable long range forces mediated by self-propelled colloidal hard spheres}

\author{Ran Ni}
\email{rannimail@gmail.com}
\affiliation{%
Van $'$t Hoff Institute for Molecular Sciences, Universiteit van Amsterdam, Science Park 904, 1098 XH Amsterdam, The Netherlands
}%
\affiliation{%
Laboratory of Physical Chemistry and Colloid Science, Wageningen University, Dreijenplein 6, 6703 HB Wageningen, The Netherlands
}%

\author{Martien A. Cohen Stuart}
\affiliation{%
Laboratory of Physical Chemistry and Colloid Science, Wageningen University, Dreijenplein 6, 6703 HB Wageningen, The Netherlands
}%

\author{Peter G. Bolhuis}
\affiliation{%
Van $'$t Hoff Institute for Molecular Sciences, Universiteit van Amsterdam, Science Park 904, 1098 XH Amsterdam, The Netherlands
}%
\begin{abstract}
Using  Brownian dynamics simulations, we systematically study the effective interaction between two parallel hard walls in a 2D suspension of self-propelled (active) colloidal hard spheres, and we find 
that the effective force between two hard walls can be tuned from a long range repulsion into a long range attraction by changing the density of active particles. At relatively high densities, the active hard spheres can form a dynamic crystalline bridge, which induces a strong oscillating long range dynamic wetting repulsion between the walls. With decreasing density, the dynamic bridge gradually breaks, and an intriguing long range dynamic depletion attraction arises. 
 A similar effect occurs in a quasi-2D suspension of self-propelled colloidal hard spheres by changing the height of the confinement. Our results open up new possibilities to manipulate the motion and assembly of microscopic objects by using active matter.
\end{abstract}

\maketitle
In past decades, the non-equilibrium dynamics of self-propelled particles has attracted an increasing amount of interest, which originates from the aim to understand the intriguing self-organization phenomena in nature like bird flocks, bacteria colonies, tissue repair, and cell cytoskeleton~\cite{revscience2012}. Very recently, breakthroughs in particle synthesis have enabled the fabrication of artificial colloidal microswimmers that show a high potential for applications in biosensing, drug delivery, etc~\cite{ebbens2010}.
A number of different active colloidal systems have been realized in experiments, such as colloids with magnetic beads acting as artificial flagella~\cite{dreyfus2005}, catalytic Janus particles~\cite{howse2007,erbe2008,palacci2010,baraban2012}, laser-heated metal-capped particles~\cite{volpe2011}, light-activated catalytic colloidal surfers~\cite{palacci2013}, and platinum-loaded stomatocytes \cite{wilson2012}. In contrast to passive colloids undergoing Brownian motion due to random thermal fluctuations of the solvent, active self-propelled colloids experience an additional force due to internal energy conversion. 
Although the long time dynamics of self-propelled particles is still Brownian, with a mean square displacement proportional to time~\cite{lowen2011}, the random self-propulsion has produced a variety of strikingly new phenomena, which were never observed in corresponding systems of passive particles~\cite{saric2013}, e.g. bacteria ratchet motors~\cite{leonado2010}, meso-scale turbulence ~\cite{wensink2012}, living crystals~\cite{palacci2013} etc. In this work, we address the question whether active matter can serve as a medium to generate unexpectedly large effective interactions between large immersed objects to direct their motion and assembly~\cite{angelani2011}. To this purpose, we employ a simple, yet representative model system, in which we study the effective interaction between two parallel hard walls immersed in suspensions of self-propelled colloidal hard spheres. 
We find that when the density of particles is relatively high, a dynamic crystalline bridge forms between the two walls, which induces a strongly oscillating repulsive dynamic wetting force, with a range depending on the size of the dynamic clusters. With decreasing density of particles, this dynamic crystalline bridge becomes smaller, and the effective force between the two walls develops a long attractive tail. 
Intriguingly, in the limit of zero density, i.e. non-interacting ideal self-propelled particles, the effective interaction turns into a long range dynamic depletion force, with a range depending on the persistence length of the mean free path of the particles, which can be tuned by varying the self-propulsion on the particle. Therefore, our results suggest a novel way to tune the interaction between large objects by immersing them in suspensions of small self-propelled colloids. The sign of interaction can be tuned from long range repulsive to long range attractive by changing the density of particles, and the range of interaction can be controlled by varying the magnitude of self-propulsion on the particles.

\begin{figure*}[ht!]
\begin{center}
\centerline{\includegraphics[width = \textwidth]{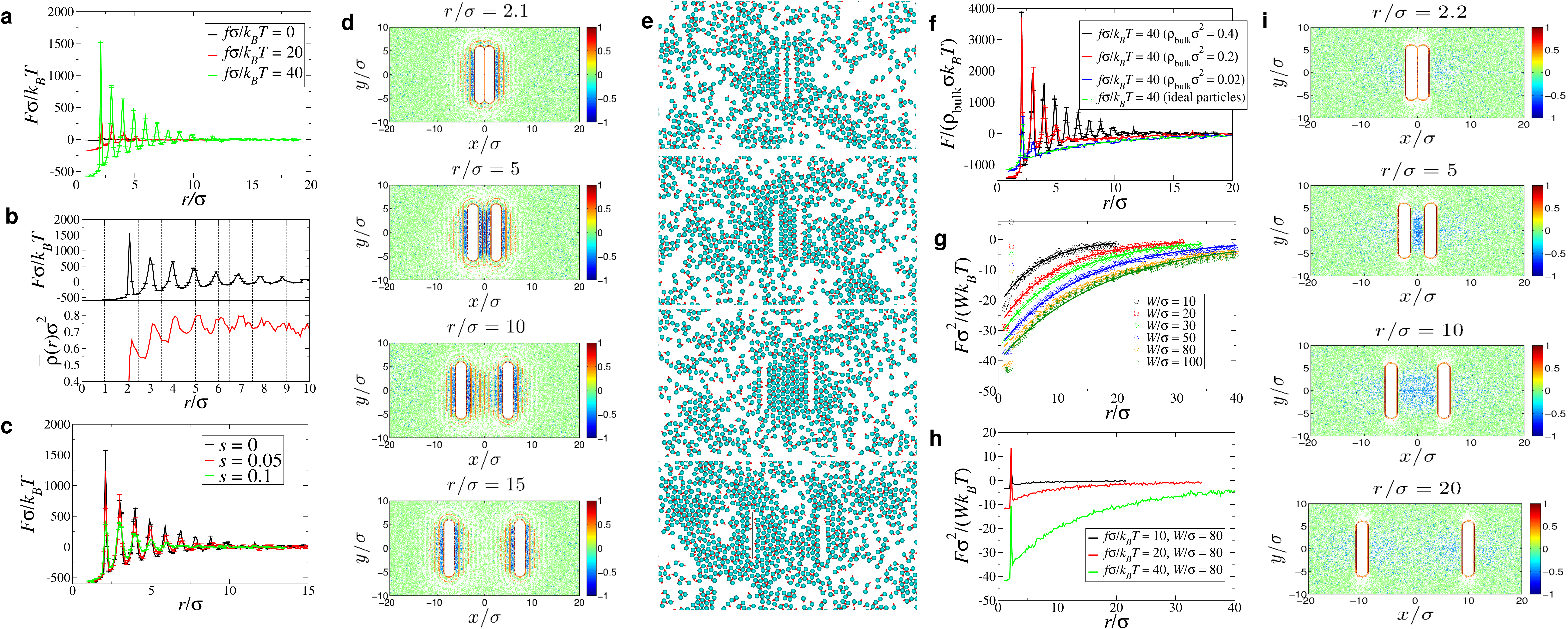}}
\caption{(color online)
(a) Effective forces $F\sigma/k_B T$ as a function of wall-to-wall distance $r$ 
at density $\rho_{bulk} \sigma^2 = 0.4$ with various activity $f$.
(b) $F\sigma/k_B T$ and the average density $\overline{\rho}(r)$ inside the confinement $-0.5r \le x \le 0.5r, -0.5W \le y \le 0.5W$ as function of $r$ 
with $f \sigma /k_B T = 40$ and $\rho_{bulk} \sigma^2 = 0.4$.
(c) $F\sigma/k_B T$ as a function of $r$ 
at $\rho_{bulk} \sigma^2 = 0.4$ with various Gaussian size polydispersity $s$.
(d) Reduced density distribution $\log_{10}[\rho(x,y)/\rho_{bulk}]$ of active particles 
 with $f \sigma /k_B T = 40$ at $\rho_{bulk} \sigma^2 = 0.4$, with a wall-to-wall distance $r/\sigma = 2.1, 5, 10$ and $20$.
(e) Typical snapshots of systems for density distributions in (d), respectively.
(f) $F/(\rho_{bulk}\sigma k_B T)$ 
for $f\sigma/k_BT = 40$ as function of wall-to-wall distance $r$ for various particle densities.
(g) $F\sigma^2/(W k_B T)$ 
for a self-propelled ideal particle system with $f\sigma/k_BT = 40$ and $\rho_{bulk} \sigma^2 = 0.2$ 
for various wall sizes $W$. The lines are fits with $\sim -\exp(-r/\xi)$.
(h) $F\sigma^2/(W k_B T)$  for ideal particles with density $\rho_{bulk} \sigma^2 = 0.2$ for various self-propulsion  and 
a wall size $W/\sigma = 80$.
(i) Reduced density distribution $\log_{10}[\rho(x,y)/\rho_{bulk}]$ for ideal particle systems with $f\sigma/k_BT = 40$ and $r/\sigma = 2.2, 5, 10$ and 20, where the wall size is $W/\sigma = 10$, and $\rho_{bulk} \sigma^2 = 0.4$.
\label{fig1}}
\end{center}
\end{figure*}

We consider a 2D suspension of $N$ (between 10,000 and 100,000) self-propelled colloidal hard spheres. The interaction between particles is modeled 
with  a steep Weeks-Chandler-Andersen (WCA) potential,
which is a good approximation of the interaction of colloidal hard spheres~\cite{supinfo,filionjcp2011,tanaka2010}. 
Even though the particles are driven and energy is continuously supplied to the system, we assume the solvent to be at an equilibrium temperature $T$.
The motion of particle $i$ with position $\mathbf{r}_i$ and orientation $\mathbf{\hat{u}}_i$ can be described via the overdamped Langevin equation 
$     \mathbf{\dot{r}}_i(t) = \frac{D_{0}}{k_B  T} \left \lbrack -\nabla_i U(t) + \mathbf{\xi}_i(t) + f \mathbf{\hat{u}}_i(t) \right \rbrack,$
where the potential energy $U$ 
is the sum of interactions between all particle pairs, and the short-time self diffusion coefficient $D_{0}$ of particle $i$ is proportional to its inverse diameter $1/\sigma$. A stochastic force  with zero mean, $\mathbf{\xi}_i(t)$, describes the  collisions with the solvent molecules, and satisfies $\langle \mathbf{\xi}_i(t)\mathbf{\xi}_j^T(t') \rangle = 2 (k_B  T)^2\mathbf{1}\delta_{ij}\delta(t-t')/D_{0}$ with $\mathbf{1}$ the identity matrix. In addition, the self-propulsion of particle $i$ is described by a  constant force $f$ in the direction $\mathbf{\hat{u}}_i(t)$ at time $t$, which undergoes free Brownian rotation~\cite{lowen2012} with a rotational diffusion coefficient assumed $D_r = 3 D_0/\sigma^2$ according to the Stokes-Einstein relation~\cite{ni2013}.

We first perform Brownian dynamics simulations of this 2D colloidal hard-sphere system at density $\rho_{bulk}\sigma^2 = 0.4$, which is much lower than the equilibrium crystallization density of hard disks~\cite{hdphasediag}, for hundreds of Brownian times $\tau_B= \sigma^2 /D_0$.  In equilibrium, i.e. $f = 0$, the system stays in a homogeneous fluid state (Ref.~\citep{supinfo} Fig.~S1a). It has been found that with increasing self-propulsion, the system undergoes a dynamic phase separation via nucleation and growth above a ``binodal'' point, and spinodal decomposition near a ``spinodal''~\cite{cates2008,catesJSMTE2011,speck2013,cates2013prl,cates2014theory,speck2014,brady2014,lowen2013,redner2013,fily2012,
cates2013epl,fily2014,cates2014softmatter,bocquet,palacci2013,lowen2013}. At density $\rho_{bulk} \sigma^2 = 0.4$, the binodal and spinodal of the dynamic phase separation are very close to each other~\cite{speck2014}. Therefore, with increasing  self-propulsion from $f\sigma/k_B T = 20 $ to 60, the average dynamic cluster size increases when approaching the spinodal of the dynamic phase separation, and at $f\sigma/k_B T = 80$, a full dynamic phase separation occurs  (Ref.~\citep{supinfo} Fig.~S1b-e).

Next, we put two parallel hard walls into the system, modelled as  line segments with length $W$ perpendicular to the horizontal ($x$) axis, and fixed at $(-r/2,0)$ and $(r/2,0)$, 
where $r$ is the center-to-center distance between the two hard walls.
The interaction between particle and wall is given by the WCA interaction using 
the minimal distance between the center of particle  and the line segment~\cite{supinfo}. 
A hard wall thus appears as a hard 2D spherocylinder with width $\sigma$ and cylindrical length $W$. We measured the effective force between walls with size $W/\sigma = 10$ immersed in the self-propelled hard sphere suspension at bulk density $\rho_{bulk}\sigma^2 = 0.4$ with various random self-propulsion $f$ (Fig.~\ref{fig1}a). We ensure that our system is large enough to 
neglect the effect of the walls on the bulk phase. 
When $f = 0$, the two walls only feel a short range depletion force ($F<0$ ) due to the colloidal particles, which terminates at around $r/\sigma \simeq 2$, where the particles start to fill the confinement area. When the random self-propulsion increases to $f\sigma/k_BT = 20$ and 40,
a strong oscillating force appears between the two walls, and with increasing $f$, both the strength and range of the force dramatically increase. The magnitude of the force reaches $F\sigma=1500 k_B T$ for $f \sigma /k_B T=40$.   To understand the origin of this giant force, we plot the reduced density distribution of particles $\rho(x,y)/\rho_{bulk}$ with $f\sigma/k_BT = 40$ for various wall-wall separations in Fig.~\ref{fig1}d. When the wall-to-wall distance
$r/\sigma \simeq 2$, the confined area
 is filled by a layer of particles (see Fig.~\ref{fig1}e). For 
$2 \le r/\sigma \le 10$,  a bridge of several dense layers of particles forms between the walls, visible as peaks in Fig.~\ref{fig1}d. Typical configurations of the dynamic bridge in Fig.~\ref{fig1}e show that it  has crystalline order, in which most of the particles are on a hexagonal crystalline lattice. When the wall-wall separation further increases to $r/\sigma = 15$, the dynamic bridge breaks (Fig.~\ref{fig1}e), and the interaction between the two walls vanishes. To further investigate the relationship between the effective force and the dynamic crystal layering between the two walls, we plot the average particle density inside the confinement, $\overline{\rho}(r)$, as a function of wall-wall separation $r$ in Fig.~\ref{fig1}b. For $r/\sigma < 2$, no particles are within the confinement, and the strong forces on the outside push  the two walls together. For $r/\sigma>2$, $\overline{\rho}(r)$ first increases and then starts oscillating as a function of $r$. Interestingly, the oscillation of $\overline{\rho}(r)$ strongly correlates with the oscillation of effective force between the two walls. Repulsive force, $F>0$, is always coupled with an increase of $\overline{\rho}(r)$, i.e. $\partial \overline{\rho}(r)/\partial r > 0$, while attraction, $F<0$, is always coupled with the decrease of $\overline{\rho}(r)$, i.e. $\partial \overline{\rho}(r)/\partial r < 0$. This strongly suggests that the effective force between the walls is induced by dynamic wetting of the living crystalline clusters in the system.  However, different from the attractive wetting force in equilibrium~\cite{gnan2012}, the dynamic wetting force we observed here appears to be overall repulsive, as the repulsive part is larger than the attractive part in the effective interaction. 
We also investigated the effect of polydispersity on the dynamic wetting force. As shown in Fig.~\ref{fig1}c, with increasing the size polydispersity, the dynamic wetting force becomes weaker,  due to the suppressed wetting by the polydispsersity. 
This is similar to the equilibrium wetting of hard-sphere crystals on a hard wall depending on the symmetry of the crystal and  surface properties~\cite{lowenprl2000,dijkstra2004}. We also confirm that polydispersity in the magnitude of the active force does not influence the dynamic wetting force significantly~\cite{supinfo}.

Furthermore, we study the influence of the bulk phase density of the particles $\rho_{bulk}$  on the effective wall-wall  interaction. The measured reduced forces $F/(\rho_{bulk} \sigma k_B T)$ for various $\rho_{bulk}$ are shown in Fig.~\ref{fig1}f. Decreasing the bulk density from $\rho_{bulk}\sigma^2 = 0.4$ to 0.2, reduces both the force amplitude of oscillation and its range, because the size of dynamic clusters decreases with lower density~\cite{lowen2013,redner2013,fily2012}. Surprisingly, an attractive tail starts to appear in the effective force, 
which becomes (relatively) more dominant with decreasing $\rho_{bulk}$. In dilute systems, i.e. $\rho_{bulk} \sigma^2 = 0.02$, 
except for a repulsive peak at $r/\sigma \simeq 2$, the force between the walls has become almost entirely attractive, with a contact value of $\approx -20k_BT/\sigma$ and a range of $\approx 20\sigma$. To understand this intriguing attraction, we study the  wall-wall interaction
in systems containing non-interacting ideal self-propelled particles, in which the particles do not interact with each other, but only with  the two hard walls. The density distribution for various $r$  in Fig.~\ref{fig1}i shows a strong peak when $r/\sigma \simeq 2$ due to dynamic wetting of the self-propelled particles~\cite{eplgompper2014a,eplgompper2014b}. Increasing the wall-wall separation to $r/\sigma = 5$, the dynamically wetted layers of ideal particles on the walls persist, while inside the confinement the density of particles is much lower than $\rho_{bulk}$.
Therefore, the inside surface of the walls 
is in contact with 
a lower density fluid than the outside, yielding a net attractive force pushing the two walls towards each other. As shown in Fig.~\ref{fig1}i, further increasing $r$ eliminates the density difference between the outside and inside of the confinement, and at sufficiently large  $r$ 
the attraction vanishes. 
We also plot the reduced pressure $F\sigma^2/(W k_B T)$ on the two walls as a function of $r$ 
for various wall sizes in Fig.~\ref{fig1}g.  With larger wall size $W$, the magnitude of pressure increases, but when $W/\sigma \ge 80$, the effect of the wall edges has become negligible. Moreover, the reduced pressure between walls can be well fitted with an exponential form $\sim -\exp(-r/\xi)$, where $\xi$ can be regarded as the range of the force. 
With increasing random self-propulsion, both the strength and range of the effective attraction increase.  

\begin{figure}
\begin{center}
\centerline{\includegraphics[width = 8.7cm]{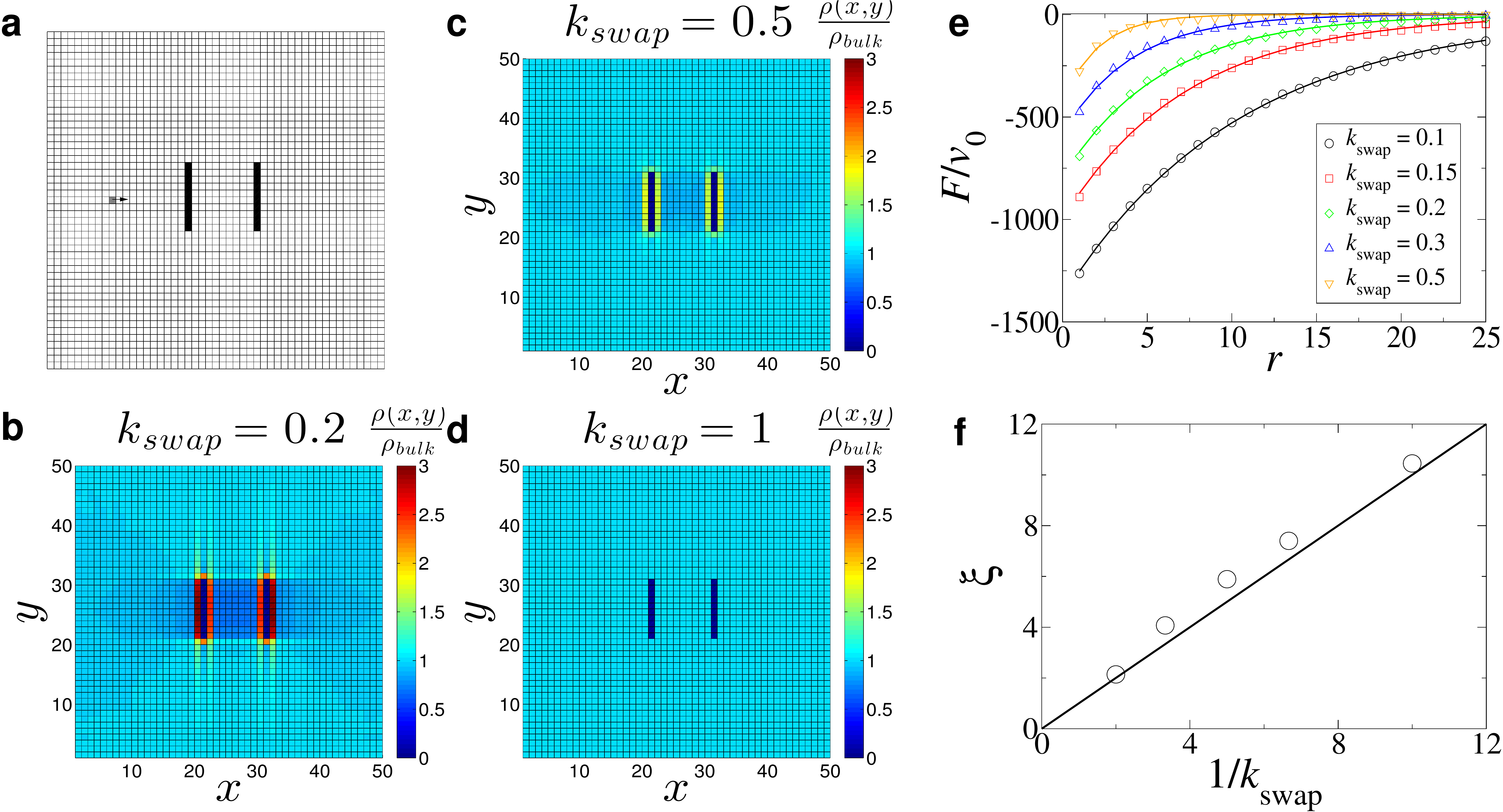}}
\caption{(color online)
(a) Simple model with one ideal particle (grey square) moving in a 2D square lattice, with two parallel impenetrable walls (dark squares).
(b-d) Reduced density distributions $\rho(x,y)/\rho_{bulk}$  
for  $k_{swap} = 0.2, 0.5$ and 1, respectively. (e) Effective force between two walls  $F/v_0$
for various $k_{swap}$ as a function of $r$.  The solid lines are the fits with $\sim -\exp(-r/\xi)$. (f) Fitted interaction range $\xi$ in (e) as a function of $1/k_{swap}$.
\label{fig2}}
\end{center}
\end{figure}

To understand the physics of the intriguing attraction mediated by self-propelled ideal particles, we studied an even simpler active particle model in which a single particle moves on a 2D ($100 \times 100$) square lattice with a constant speed of $v_0=1$ lattice site per step in the $\pm x$ or $\pm y$ direction  (see Fig.~\ref{fig2}a).
The particle has a probability $k_{swap}$ to rotate by $\pm 90$ degrees. 
In this model, the only controlling parameter is $k_{swap}$. When $k_{swap} < 1$, at each step the particle has a probability of $1-k_{swap}$ to keep its current direction. Therefore, $k_{swap}$ characterizes the persistence length in the random walk $l_p$; this is the expected path length before the particle turns, $l_p = \sum_{i=1}^\infty  i k_{swap}  (1-k_{swap})^{i-1} = 1/k_{swap}$.
Placing two parallel hard walls of 10 sites each, with the rule that  
if the particle hits a wall it remains at its lattice position, we perform Monte Carlo simulations long enough to reach a steady state ($\sim 10^8$ steps).

Figure~\ref{fig2}b-d shows, for a fixed wall-wall separation 
the reduced probability distributions of the particle on the square lattice with various $k_{swap}$. 
When $k_{swap} = 1$, this  distribution is homogeneous, as expected. However, at $k_{swap} = 0.2$ or 0.5, the distribution becomes spatially heterogeneous, and the probability density inside the confinement is lower than outside. 
The force on the walls follows from the average probability of finding particles colliding with the wall (see  Fig.~\ref{fig2}e). 
The decay constant $\xi$ (obtained from fitting to $F \sim -\exp(-r/\xi)$) depends on  $k_{swap}$ as $\xi \sim 1/k_{swap}$ (see  Fig.~\ref{fig2}f), as the range of the non-equilibrium dynamic depletion force is close to the persistence length, which is $l_p=1/k_{swap}$. 
For  active Brownian hard spheres, this persistence length, signifing the loss of correlation of the direction of active motion, is not a well-defined single quantity due to the thermal fluctuations. However, one can still define a persistence length 
 $l_{p}^{BD}$ as the displacement at time 
$\tau_{R} = 1/D_r$ at which the orientation correlation of the active force decays by a factor $e$. 
This yields $l_{p}^{BD}/\sigma = 2.88, 5.21$ and 10.14 for $f\sigma/k_B T = 10, 20$ and 40, respectively~\cite{lowen2011}. By fitting the force with $\sim -\exp(-r/\xi)$ in Fig.~\ref{fig1}h, we obtain $\xi/\sigma = 7.41, 12.03$ and 16.774 for $f\sigma/k_B T = 10, 20$ and 40, respectively. 
Thus, $l_p^{BD}$ and $\xi$ are both on the same scale, and increase with increasing the self-propulsion. This suggests that our model of a single active lattice particle can capture the essential physics of the active Brownian particle. Nevertheless, we
stress that the quantitative physics of the novel dynamic depletion induced by active particles, even in the simple lattice model
requires further studies.

\begin{figure}[t!]
\begin{center}
\centerline{\includegraphics[width = 8.7cm]{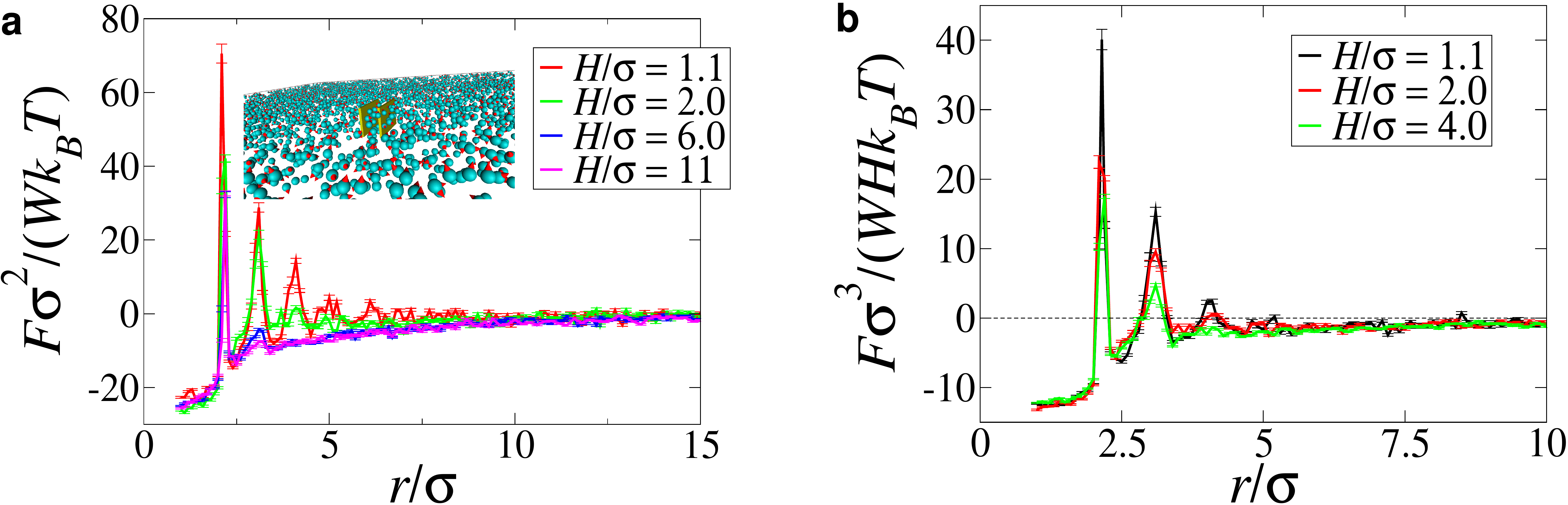}}
\caption{ (color online)
(a) Main: $F\sigma^2/(Wk_B T)$ as function of wall-wall separation $r/\sigma$ for 
$f\sigma/k_BT = 40$ in quasi-2D confinement with varying height $H/\sigma$, at fixed 2D  number density $N \sigma^2/A = 0.4$.
Inset: a typical system snapshot with two walls as yellow bricks. 
(b) $F\sigma^3/(WHk_B T)$ 
as function of 
$r/\sigma$ for 
$f\sigma/k_BT = 40$ 
and varying $H/\sigma$, where the 3D number density is fixed at $N \sigma^3/(A H) = 0.2$.
\label{fig3}}
\end{center}
\end{figure}

Finally, to make our findings relevant for micro-fluidic experiments, we study self-propelled colloidal hard spheres confined in a quasi-2D confinement (see inset Fig.~\ref{fig3}a). The confinement height between two horizontal hard walls is $H$~\cite{supinfo}.
Figure~\ref{fig3}a shows the effective interaction between two parallel finite vertical walls with $W/\sigma = 10$, for a  random self-propulsion  $f\sigma/k_BT = 40$, and  a fixed 2D  bulk number density $N\sigma^2/A =0.4$, where $A$ is the area in the $x,y$ plane. 
With increasing confinement height $H$, both the range and strength of the oscillating force mediated by confining dynamic clusters between the two vertical walls decrease, 
as these clusters 
shrink with larger $H$.
A long range attraction starts to dominate the effective interaction 
similar to the case of 
 2D active hard spheres. 
When $H/\sigma \simeq 6$, this long range attraction has a contact value around $-200k_B T /\sigma$.  Thus, by changing the confinement height, one can tune the effective interaction mediated by active hard spheres from a long rang repulsion into a long range attraction.
Figure~\ref{fig3}b shows  that 
for a fixed 3D number density $N \sigma^3/(A H) = 0.2$,
  with increasing the confinement height, the oscillating repulsion in the effective interaction  decreases, and the attraction tail gradually increases.  This is similar to decreasing the density of the particles, as 
in a quasi-2D system of constant volume fraction, a larger confinement height produces more free volume for particles, which effectively decreases the ``density'' of the particles.

We stress that the activity studied here lies in realizable range in recent experiments. For instance, the light-activated colloids \cite{palacci2013}, the catalytic Janus particles \cite{baraban2012}, and the particles with an artificial magnetic flagella \cite{dreyfus2005} are capable of producing self-propulsions as high as $f \sigma/k_B T \simeq 20$, 50, 80, respectively, which should be able to produce giant effective force to drive the self-assembly of passive particle systems.
While we fix the two parallel walls during the simulation, in reality, all objects including the passive objects in the suspensions may move, so that the effect induced by active particles might be different~\cite{leonado2010}. However, it is known that the diffusivity of passive objects in solution drops dramatically with increasing sizes. Therefore, our results suggest a novel way of producing tunable long range effective interactions between colloidal particles, whose size is of orders of magnitude larger than that of the active colloidal particles, and direct their self-assembly, which was difficult to achieve by using passive particles.
In this work we focus on the effective interaction produced by the general non-equilibrium feature of active matter, and neglected hydrodynamic effects that 
would be interesting for future experimental and theoretical studies ~\cite{Yamamoto2013}. Finally, 
in the previous experiment studying the effective interaction between colloidal spheres mediated by active matter, i.e. bacteria, different from this work, only a short range repulsion instead of any long range forces was observed~\cite{angelani2011}. The reason is that the shape of the active and passive colloids also plays an important role~\cite{harder2014}.

\begin{acknowledgments}
After completing this work, we became aware of a very recent work by Ray {\em et al.} showing that in another model of  run-and-tumble active particles, a similar dynamic depletion force was also observed, which suggests that the dynamic depletion force may generally appear in many active matter systems~\cite{ray2014}.
R.N. and M.A.C.S acknowledge financial support from the European Research Council through Advanced Grant 267254 (BioMate).
This work is part of the research programme VICI 700.58.442, which is financed by the Netherlands Organization for Scientific Research (NWO).
 R.N. has been supported by an NWO VENI grant.
\end{acknowledgments}

\bibliography{ref}

\clearpage
\newcommand{\beginsupplement}{%
        \setcounter{table}{0}
        \renewcommand{\thetable}{S\arabic{table}}%
        \setcounter{figure}{0}
        \renewcommand{\thefigure}{S\arabic{figure}}%
     }
\onecolumngrid
\beginsupplement
\section{Supplementary Materials}
In the 2D system of active hard spheres, the interaction between particle $i$ and $j$ is modeled
with a steep Weeks-Chandler-Andersen (WCA) potential given by
\begin{equation}
        \frac{U_{WCA}(r_{ij})}{k_B T} = \left\{
        \begin{array}{lc}
        4\epsilon \left[ \left( \frac{\sigma}{r_{ij}}\right)^{12} - \left( \frac{\sigma}r_{ij}\right)^{6} + \frac{1}{4}\right] & r_{ij}/\sigma \le 2^{1/6},\\
        0 & r_{ij}/\sigma > 2^{1/6},            \end{array}
        \right.
\end{equation}
where $r_{ij}$ is the center-to-center distance between the two spheres, and $\sigma$ is the diameter of the particles, with $k_B$ and $T$ the Boltzmann constant and the temperature of the system
, respectively. Here, the strength of the interaction is set $\epsilon = 40$. Moreover, the interaction between particle $i$ and wall $j$ is given by $U_{wall}(i,j) = U_{WCA}(r_{ij})$, in which $
r_{ij}$ is minimal distance between the center of particle $i$ and points on the line segment $j$.

In the quasi-2D system of active hard spheres confined betwen two horizontal walls, the interaction between particle $i$ at $\mathbf{r}_i = (x_i,y_i,z_i)$ and the two horizontal walls is
\begin{equation}
        \frac{U_{conf}(\mathbf{r}_i)}{k_B T} = \left\{
        \begin{array}{lc}
        0 & 0.5\sigma \le z_i \le H - 0.5\sigma,\\
        \infty & otherwise,     
        \end{array}
        \right.\end{equation}
where $H$ is the height of the confinement.

 \begin{figure}[H]
 \centerline{\includegraphics[width = 0.9 \textwidth]{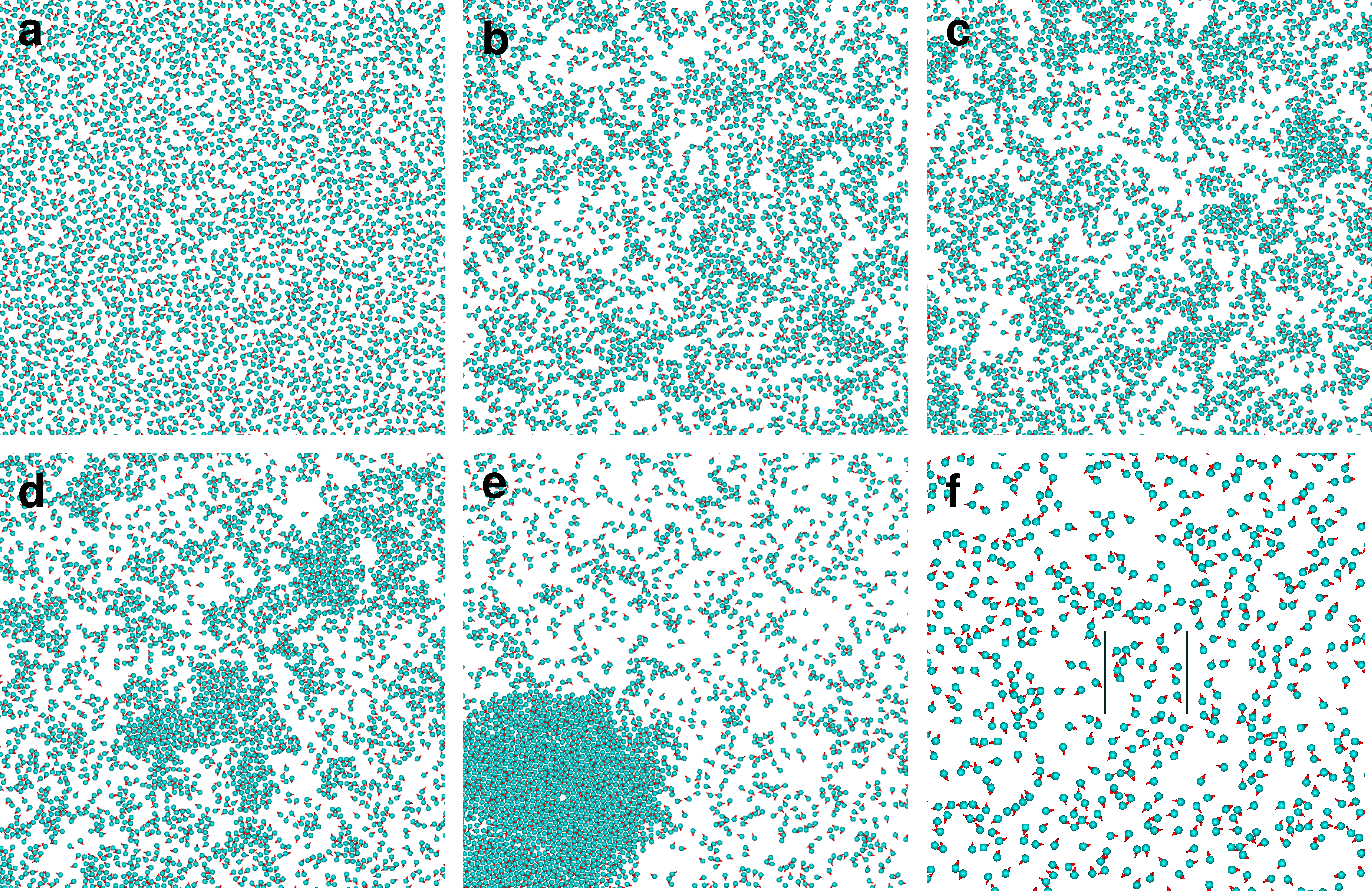}}
 \caption{
 (a-e) Typical snapshots of 2D systems of self-propelled colloidal hard spheres at density $\rho_{bulk} \sigma^2= 0.4$ with self-propulsion $f\sigma/k_B T = 0, 20, 40, 60$ and 80, respectively.
 (f) Illustration of the system to study the interaction mediated by the dynamic clusters of self-propelled colloidal hard spheres, and the two vertical lines indicate the two parallel hard walls. The red arrows indicate the direction of random self-propulsion.
 \label{figS1}}
\end{figure}

\begin{figure}[H]
 \centerline{\includegraphics[width = 0.45 \textwidth]{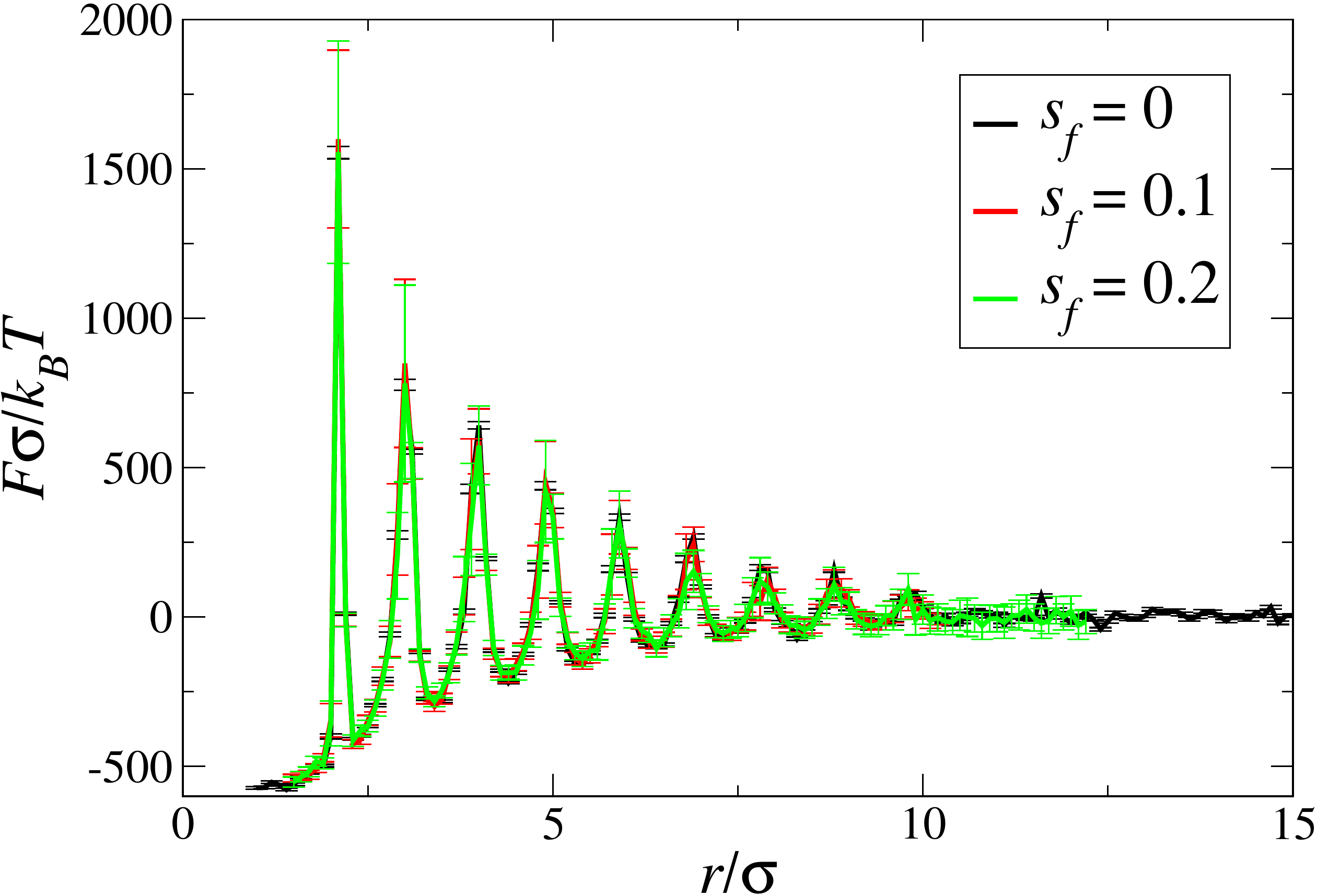}}
 \caption{
  Effective interaction between two parallel walls $F\sigma/k_BT$ as a function of wall-wall distance $r$ for systems of active hard spheres with different polydisperse active propulsions $s_f$. The density of the system is fixed at $\rho_{bulk} \sigma^2 = 0.4$, and the average self-propulsion is $f\sigma/k_B T = 40$ with $\sigma$ the particle diameter.
 \label{figS2}}
\end{figure}

The effect of polydispersity of active force on the dynamic wetting force is shown in Fig.~\ref{figS2}. The self-propulsion on particle $i$ is given by
\begin{equation}
        f_{i}  = f \cdot \left[1+\xi(s_f) \right],
\end{equation}
where $f$ is average active force in the system, and $\xi(s_f)$ is a Gaussian random number with zero mean and the standard deviation $s_f$.

\end{document}